\documentclass[reprint, aps, prx, amsmath, amssymb, longbibliography, superscriptaddress]{revtex4-2}

\usepackage{graphicx}
\usepackage{mathrsfs,esint}
\usepackage{relsize}
\usepackage{physics}
\usepackage{dsfont}
\usepackage{svg}

\usepackage[caption=false]{subfig}
\usepackage[percent]{overpic}

\usepackage[bookmarks=true,
   colorlinks=true,
   linkcolor=blue,
   urlcolor=blue,
   citecolor=blue,
   bookmarks=true,
   hyperindex=true
]{hyperref}

\newcommand{\Zn}{\mathbb{Z}_n}
\newcommand{\Zthree}{\mathbb{Z}_3}
\newcommand{\Ztwo}{\mathbb{Z}_2}

\renewcommand{\Re}{\operatorname{Re}}
\renewcommand{\Im}{\operatorname{Im}}

\begin{document}

\title{Cat states in one- and two-mode \texorpdfstring{$\Zthree$}{Z\_3} Rabi models}

\author{Anatoliy I. Lotkov}
\affiliation{Department of Physics, University of Basel, Klingelbergstrasse 81, CH-4056 Basel, Switzerland}

\author{Denis V. Kurlov}
\affiliation{Department of Physics, University of Basel, Klingelbergstrasse 81, CH-4056 Basel, Switzerland}

\author{Valerii K. Kozin}
\affiliation{Department of Physics, University of Basel, Klingelbergstrasse 81, CH-4056 Basel, Switzerland}

\author{Jelena Klinovaja}
\affiliation{Department of Physics, University of Basel, Klingelbergstrasse 81, CH-4056 Basel, Switzerland}

\author{Daniel Loss}
\affiliation{Department of Physics, University of Basel, Klingelbergstrasse 81, CH-4056 Basel, Switzerland}


\begin{abstract}
We investigate one- and two-mode variants of the $\mathbb{Z}_3$-symmetric quantum Rabi model, which describe the interaction of a qutrit with one or two bosonic modes and are directly relevant for circuit-QED and spin-qudit platforms. We find a canonical transformation that allows one to obtain the spectrum of the $\Zthree$ Rabi model using perturbation theory in a magnetic field. We show that in a certain parameter regime (deep-strong–coupling and a small magnetic field) the three lowest eigenstates become $\mathbb{Z}_3$ qutrit-boson cat states. In order to characterize these states we introduce a joint qutrit–boson Wigner function and derive its closed-form expression for the qutrit-boson cat states. Numerical calculations across a wide range of coupling strengths show that the proposed Wigner function is a useful tool that allows one to unambiguously identify the ${\mathbb Z}_{3}$ qutrit-boson cat states.
\end{abstract}

\textbf{}\maketitle

\section{Introduction}

Quantum systems with $\Zn$ symmetry have attracted significant attention in various areas of physics. They usually appear as a natural generalization of more familiar $\Ztwo$-symmetric systems. For example, in condensed matter physics, the $\Zn$ Potts model generalizes one of the simplest integrable systems, the Ising model \cite{wu_potts_1982,fortuin_randomcluster_1972,temperley_relations_1971,baxter_new_1988,lotkov_floquet_2022}. Similarly, the $\Zn$ symmetry arises in the context of topological phases of matter. There, the $\Zn$-symmetric parafermionic edge states are a generalization of the Majorana edge states~\cite{fendley_free_2013,fendley_parafermionic_2012, teixeira_edge_2022,alicea_topological_2016,cobanera_fock_2014,kurlov_mesoscale_2024, hutter_quantum_2016,klinovaja_parafermions_2014}.
Furthermore, $\Zn$ symmetry plays an important role in quantum technologies. Indeed, the qudit‑based quantum processors extend the usual qubit paradigm~\cite{kiktenko_colloquim_2025,wang_qudits_2020, kiktenko_single_2015, kiktenko_scalable_2020,proctor_quantum_2019,farinholt_ideal_2014,brylinski_universal_2001}. Specifically, leading platforms for quantum computations, such as trapped ions, superconducting qubits, or semiconducting spin qubits, inherently possess more than two energy levels that can be used for computations, so that using $n$-level qudits (with their intrinsic $\Zn$ symmetry) is a straightforward and powerful extension of qubit-based quantum computing. 
Finally, in quantum optics, the paradigmatic Rabi model, which possesses $\Ztwo$ symmetry and describes the light-matter interaction \cite{braumuller_analog_2017, braak_integrability_2011,hwang_quantum_2015, chen_shortcuts_2021}, can be also extended to the $\Zn$-symmetric case \cite{albert_quantum_2012, zhang_z_n_2014, sedov_chiral_2020,kozin_quantum_2024}. 

In this paper we consider the $\Zn$ Rabi model. For the sake of simplicity, we restrict ourselves to the $\Zthree$-symmetric case. Although our results can be readily extended to $\Zn$ symmetry with arbitrary integer $n > 1$, the simplest non-trivial case of the $\Zthree$ Rabi model already exhibits the phenomena that we are interested in, while allowing for more transparent treatment. We discuss in detail several variants of the $\Zthree$ Rabi model, highlighting their similarities and differences. 
Our main focus is on a $\Zthree$ Rabi model with two boson modes because its optomechanical implementation was proposed~\cite{sedov_chiral_2020}. In a companion paper~\cite{lotkov_superconducting_}, we show that this particular model can be simulated by a superconducting circuit. Here, we demonstrate that the $\Zthree$ Rabi model exhibits behavior similar to the superradiant phase found in the ordinary Rabi model. 
Specifically, in the regime of deep-strong light-matter coupling and weak magnetic field, the ground state becomes a $\Zthree$-symmetric generalization of a cat state. 

Recently, there has been a renewed interest in cat states as they find applications in quantum error correction (QEC) codes~\cite{vlastakis_deterministically_2013,malbouisson_highergeneration_1999,bergmann_quantum_2016,grimm_stabilization_2020}. Cat states in a Kerr resonator with a two-photon dissipation were even proposed as a platform for fault-tolerant quantum computation \cite{guillaud_repetition_2019}. While the cat-state QEC has mostly focused on qubits, a similar construction also works for qudits. In this work, we obtain $\Zthree$ cat states in a unitary system that does not require an engineered dissipation, thus simplifying possible experimental realizations.  

In extensive research on cat states in various areas~\cite{gou_vibrational_1996,kira_quantum_2011,bild_schrodinger_2023}, the Wigner function has proven to be to be an extremely useful and convenient tool. While both boson and qudit Wigner functions have been studied, in this work we introduce a joint qudit-boson Wigner function. Applying it to the $\Zthree$ Rabi model, we show how the $\Zthree$ cat states appear in the deep-strong coupling regime. More generally, we believe that the joint qudit-boson Wigner functions offer a powerful method to explore light-matter interaction and hybrid systems.

The rest of the paper is organized as follows. In Sec.~\ref{sec:1-mode-rabi}, we briefly discuss the general case of the $\Zn$ Rabi model, introduce the one-mode $\Zthree$ Rabi model, and study its properties. Then, in Sec.~\ref{sec:2-mode-rabi}, the two-mode $\Zthree$ Rabi model is studied in the same way. Next, in Sec.~\ref{sec:wigner-function}, we introduce the joint qutrit-boson Wigner functions and use them to study the eigenstates of the $\Zthree$ Rabi model found in the previous section. Finally, in Sec.~\ref{sec:conclusions}, we conclude. Some technical details are deferred to Apps.~\ref{su2-symmetry}-\ref{app:equivalent_two_mode_Rabi_model}.

\section{One-mode Rabi model} \label{sec:1-mode-rabi}

In this section, we first briefly discuss the general case of the $\Zn$ Rabi model for the sake of completeness. Then, we focus on the special case of $\Zthree$ symmetry. We show that in the regime of deep-strong coupling and weak magnetic field the lowest eigenstates of the $\Zthree$ Rabi model correspond to the $\Zthree$ cat states. 
In order to demonstrate this, we employ a novel canonical transformation that allows for a transparent perturbative treatment in the relevant regime.

\subsection{\texorpdfstring{$\Zn$}{Z\_n}-symmetric case}

The one-mode $\mathbb{Z}_n$ Rabi model describes the interaction between an $n$-level qudit and a boson mode~\cite{albert_quantum_2012, zhang_z_n_2014}. The Hamiltonian reads
\begin{equation}
\label{Zn-Rabi-Hamiltonian}
   H = \Omega  a^{\dagger}  a + \sum_{j=1}^{n-1} B_j  Z^j - \lambda ( a  X^{\dagger} +  a^{\dagger} X),
\end{equation}
where $a$ ($a^{\dag}$) is a canonical boson annihilation (creation) operator, $X$ and $Z$ are the generalized Pauli matrices (commonly referred to as the shift and clock matrices, respectively), $\Omega$ is the boson mode frequency, $\lambda$ is the boson-qudit coupling strength, $B_j$ are parameters that define the qudit energy levels, and the integer $n$ is larger than one. To ensure Hermiticity, we require $B_j = B_{n-j}^*$. The $n \times n$ clock and shift matrices $X$ and $Z$ obey the following algebraic relation:
\begin{equation}
    Z^3 = 1, \quad X^3 = 1, \quad ZX = \omega X Z,
\end{equation}
where $\omega = \exp[2\pi i/n]$ is the $n$th principle root of unity.
The $X$ and $Z$ matrices are unitary and they generalize the Pauli matrices $\sigma_x$ and $\sigma_z$, correspondingly. However, they are not Hermitian. Hereafter, we call the purely qudit term $\sum_j B_j Z^j$ a ``magnetic'' term, since its $\Ztwo$ analogue is the Zeeman term.

We also define two bases in the qudit space: the computational basis diagonalizing the $Z$ matrix $Z|j\rangle = \omega^j |j\rangle,\, j=0,\,\dots, \,n-1$, and the Fourier basis diagonalizing the $X$ matrix $X|\omega^k\rangle = \omega^{-k} |\omega^k\rangle,\, k=0,\,\dots, \,n-1$. These two bases are related to each other by the discrete Fourier transform:
\begin{equation}
    |\omega^k\rangle = \frac{1}{\sqrt{n}}\sum\limits_{j=0}^{n-1} \omega^{kj}|j\rangle.
\end{equation}
In the computational basis, the $X$ and $Z$ matrices have the following form:
\begin{equation}
    X_{ij} = \delta_{(i-j-1\operatorname{mod}n),\, 0}, \quad Z_{ij} = \delta_{ij} \omega^{j-1}.
  \end{equation}

The model has the $\mathbb{Z}_n$ symmetry, which counts the number of boson modes and qudit excitations modulo $n$. One can think of the system as if there is an excitation hopping between boson and qudit degrees of freedom. The explicit form of the $\mathbb{Z}_n$ symmetry generators is
\begin{equation}
\label{Zn-symmetry}
     {\cal P} = \exp\left[\frac{2\pi i}{n}  a^{\dagger}  a\right] \, Z.
\end{equation}

More general $\Zn$-symmetric variants of Eq.~\eqref{Zn-Rabi-Hamiltonian} are possible. In particular, terms like $a^2 (X^{\dagger})^2, \, ( a^{\dagger})^2 X^2$  are also allowed since they conserve the total number of excitations modulo $n$. However, these terms make the system more complicated, and we do not consider them in this paper.

\subsection{\texorpdfstring{$\Zthree$}{Z\_3}-symmetric case}

For simplicity, from now on we restrict ourselves to the case of the $\Zthree$ Rabi model, although most of our results can be straightforwardly generalized to $\Zn$ symmetry with arbitrary integer $n > 1$. 
The Hamiltonian~(\ref{Zn-Rabi-Hamiltonian}) then reduces to
\begin{equation}
\label{z3-Rabi-Hamiltonian}
     H_{\text{R}1} = \Omega  a^{\dagger}  a + B (e^{i\phi} Z + e^{-i\phi} Z^{\dagger}) - \lambda ( a^{\dagger} X +  a X^{\dagger}),
\end{equation}
where we defined $B_1 = B e^{i\phi},\, B_2 = B e^{-i\phi}$, with $B$ and $\phi$ being the absolute value and phase. The subscript ``R1'' on the Hamiltonian~(\ref{z3-Rabi-Hamiltonian}) emphasizes that it describes the one-mode $\Zthree$ Rabi model, distinguishing it from the two-mode $\Zthree$ Rabi model discussed below.
In the $\Zthree$ case the $X$ and $Z$ matrices read
\begin{equation}
    X = \begin{pmatrix} 
         0 & 0 & 1 \\
         1 & 0 & 0 \\
         0 & 1 & 0 
        \end{pmatrix}, \quad
    Z = \begin{pmatrix}
         1 & 0 & 0 \\
         0 & \omega & 0 \\
         0 & 0 & \omega^2
        \end{pmatrix}.
\end{equation}
Thus, the Hamiltonian~(\ref{z3-Rabi-Hamiltonian}) can be written as a $3\times 3$ matrix:
\begin{equation}
     H_{\text{R}1} = \begin{pmatrix}
    \Omega  a^{\dagger}  a + \epsilon_0 & -\lambda  a^{\dagger} & -\lambda  a \\
    -\lambda  a & \Omega  a^{\dagger}  a + \epsilon_1 & - \lambda  a^{\dagger} \\
    -\lambda  a^{\dagger} & -\lambda  a &\Omega  a^{\dagger}  a + \epsilon_2 
    \end{pmatrix},
\end{equation}
where $\epsilon_k = 2 B \cos(\phi + 2\pi k/3)$, with $k=0,1,2.$ 
The $\Zthree$ symmetry generator follows directly from Eq.~(\ref{Zn-symmetry}):
\begin{equation}
\label{z3-symmetry}
     {\cal P}_{\text{R}1} = \exp\left[\frac{2\pi i}{3}  a^{\dagger}  a\right] Z.
\end{equation}

We are interested in obtaining the ground state of the Hamiltonian~(\ref{z3-Rabi-Hamiltonian}). In the next subsection, we describe a canonical transformation that simplifies the perturbation theory in the regime of deep-strong coupling $\lambda \gtrsim \Omega$ and weak magnetic field $\Omega \gg B$.

\subsection{Canonical transformation}

Different ways exist to demonstrate the cat-state structure of the ground state in the (ordinary) Rabi model. For instance, using a variational method \cite{zhang_analytical_2011} or the Schrieffer-Wolff transformation \cite{hwang_quantum_2015}. However, we find an approach based on a canonical transformation that decouples qubit degrees of freedom~\cite{benivegna_structure_1987} to be the most straightforward.  
This approach can be extended to the $\Zthree$ Rabi model, as we demonstrate in this section.
We introduce a canonical transformation that decouples qutrit degrees of freedom from the boson ones:
\begin{equation}
\label{canonical-1-mode}
    b =  a X^{\dagger}, \qquad  b^{\dagger} =  a^{\dagger} X. 
\end{equation}
Intuitively, the operator $b$ annihilates a boson while simultaneously lowering the qutrit state. The newly-introduced boson creation/annihilation operators satisfy the canonical commutation relations:
\begin{equation}
    [ b, \,  b^{\dagger}] = X[ a, \,  a^{\dagger}]X^{\dagger} = 1.
\end{equation}
However, they no longer commute with qutrit operators. Then, we eliminate the qutrit degrees of freedom altogether. First of all, the harmonic oscillator term and the interaction term can be easily expressed in terms of the new operators:
\begin{equation}
\label{shifted-harmonic-oscillator}
    \begin{aligned}
&\Omega  a^{\dagger}  a - \lambda( a^{\dagger} X +  a X^{\dagger})  \\
& = \Omega ( b^{\dagger} - \lambda/\Omega)( b - \lambda/\Omega) - \lambda^2/\Omega,
\end{aligned}
\end{equation}
which is simply a shifted harmonic oscillator.

However, we are still left with the ``magnetic'' term $B(e^{i\phi} Z + e^{-i\phi} 
Z^{\dagger})$.  It can also be expressed in terms of the operators $b$ and $b^{\dagger}$. To do so, we decompose the full Hilbert space into three subspaces with a fixed $\Zthree$ parity $\mathcal{H} = \mathcal{H}_0 \oplus \mathcal{H}_1 \oplus \mathcal{H}_2$. Here $\mathcal{H}_k$ is an eigenspace of ${\cal P}_{\text{R1}}$  [see Eq.~(\ref{z3-symmetry})] with a fixed eigenvalue:
\begin{equation}
\label{z3-restriction}
    {\cal P}_{\text{R1}} \big |_{\mathcal{H}_k} = \omega^k .
\end{equation}
Here we used the vertical line to denote the restriction of the operator to the subspace, i.e, for an operator $\mathcal{O}:\: \mathcal{H} \to \mathcal{H}$ and a linear subspace $\mathcal{H}' \subset \mathcal{H}$, we define $ \mathcal{O}\big |_{\mathcal{H'}}$ to be a part of the operator $\mathcal{O}$ mapping $\mathcal{H}'$ to $\mathcal{H}'$.

After substituting the definition of the $\Zthree$-symmetry generator ${\cal P}_{\text{R1}}$ into Eq.~\eqref{z3-restriction}, we can express the qutrit clock matrix as 
\begin{equation} \label{Z_subspace_restriction}
       Z \big|_{\mathcal{H}_k} = \omega^{k} \exp\left(- \frac{2\pi i}{3}  b^{\dagger}  b\right)\biggr|_{\mathcal{H}_k}.
\end{equation}
Note that the subspace index $k$ only enters the scalar phase factor $\omega^k$.
Using the relation~(\ref{Z_subspace_restriction}) one can now express the ``magnetic'' term in the Hamiltonian~(\ref{z3-Rabi-Hamiltonian}) using the new boson ladder operators: 
\begin{equation}
\label{magnetic-term-tranform}
    (e^{i\phi}Z + e^{-i\phi} Z^{\dagger})\big|_{\mathcal{H}_k} = 2 \cos\left[\frac{2\pi}{ 3}( b^{\dagger}  b - k) - \phi\right]\biggr|_{\mathcal{H}_k}.
\end{equation}
Finally, we can write the full $\Zthree$ Rabi model Hamiltonian in terms of the operators $b$ and $b^{\dagger}$: 
\begin{equation}
\label{simplified-z3-Rabi-Hamiltonian}
\begin{aligned}
     H_{\text{R}1} \big|_{\mathcal{H}_k} &= \biggl\{\Omega\left( b^{\dagger} - \lambda/\Omega\right)\left( b - \lambda/\Omega\right) - \lambda^2/\Omega \\
    & +2 B \cos\left[\frac{2\pi}{3}( b^{\dagger}  b - k) - \phi\right]\biggr\}\biggr|_{\mathcal{H}_k}.
\end{aligned}
\end{equation}
It is now straightforward to diagonalize the transformed Hamiltonian using the perturbation theory in the limit of deep-strong coupling and weak magnetic field, as we demonstrate below.

\begin{figure}[t]
\includegraphics[width=\linewidth]{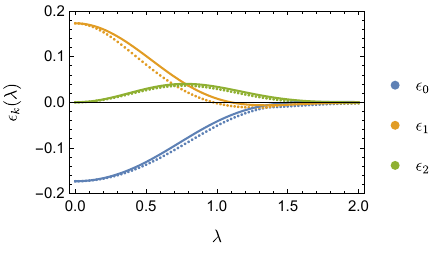}
\caption{Eigenvalues $\epsilon_k$ of the one-mode $\Zthree$ Rabi model Hamiltonian $ H_{R1}$ in Eq.~(\ref{z3-Rabi-Hamiltonian}) as functions of the qutrit-boson coupling $\lambda$ (both in units of $\Omega$). The dotted line represents the exact numerical diagonalization, while the solid line corresponds to the first-order perturbation theory~(\ref{1-mode-spectrum}). One can clearly see an excellent agreement between the numerical and the perturbative results. Parameters used: $\Omega = 1$, $B = 0.1$, $\phi = 7\pi/6$.}
\label{fig:1-mode-rabi}
\end{figure}

\subsection{Cat states in the \texorpdfstring{$\Zthree$}{Z\_3} one-mode Rabi model}
\label{1-mode-perturbation-theory}

After the canonical transformation~\eqref{canonical-1-mode}, the Hamiltonian~(\ref{simplified-z3-Rabi-Hamiltonian}) is conveniently divided into the unperturbed part and the perturbation:
\begin{equation}
    \begin{aligned}
& H_{\text{R}1} \big|_{\mathcal{H}_k} =  H_{\text{R}1}^{(0)}\big|_{\mathcal{H}_k} +  V_{\text{R}1}(k)\big|_{\mathcal{H}_k}, \\
& H_{\text{R}1}^{(0)} = \Omega( b^{\dagger} - \lambda/\Omega)( b - \lambda/\Omega) - \lambda^2/\Omega, \\
& V_{\text{R}1}(k) =  2 B \cos\left[\frac{2\pi}{3}( b^{\dagger}  b - k) - \phi\right],
\end{aligned}
\end{equation}
where the perturbation $V_{\text{R}1}(k)$ depends on the $\Zthree$ parity $k$. As we have already mentioned, the unperturbed Hamiltonian~$H_{\text{R}1}^{(0)}$ is just a shifted harmonic oscillator.

Its ground state is defined by the following requirement:
\begin{equation}
    ( b - \lambda/\Omega)|\psi_k^{\text{QB}}\rangle = 0,
\end{equation}
where the superscript ``QB'' emphasizes that this is a qutrit-boson state and the subscript $k$ indicates that the ground state belongs to the symmetry sector $\mathcal{H}_k$, i.e., ${\cal P}_{\text{R}1}|\psi_k^{\text{QB}}\rangle = \omega^k|\psi_k^{\text{QB}}\rangle$. As a result, the ground state $|\psi_k^{\text{QB}}\rangle$ is a coherent state of the boson $b$ and $b^{\dagger}$ with a coherent parameter $\lambda/\Omega$ \footnote{Here we use a nonstandard notation: ket-states with subscript $\hat a$ or $\hat b$ to distinguish between different bases for the boson modes. The subscript $k$ indicates the $\Zthree$ parity sector. Added a sentence about cross-sector term.}
\begin{equation}
    |\psi_k^{\text{QB}}\rangle = |\lambda/\Omega\rangle_{\hat b,k}.
\end{equation}

In terms of the original degrees of freedom, the state $|\psi_k^{\text{QB}}\rangle$ turns out to be the $\Zthree$ cat state:
\begin{equation}
\begin{aligned}
    |\psi_k^{\text{QB}}\rangle = &\frac{1}{\sqrt{3}}\left[|\lambda/\Omega\rangle_{\hat a}|\omega^0 \rangle + \omega^k |\omega \lambda/\Omega\rangle_{\hat a} |\omega\rangle \right. \\
    &+ \left. \omega^{2k} |\omega^2 \lambda/\Omega\rangle_{\hat a} |\omega^2 \rangle\right].
\end{aligned}
\end{equation}
Here, we again can see a parallel with the ordinary $\Ztwo$ Rabi model, whose ground states are the usual cat states  \cite{hwang_variational_2010, ashab_qubit_2010}.

Next, we use the perturbation $ V_{\text{R}1}$ to lift the degeneracy of the states $|\psi_k^{\text{QB}}\rangle$. Although the three states $|\psi_k^{\text{QB}}\rangle$ with $k=0,1,2$ are degenerate eigenstates of the unperturbed Hamiltonian, they still constitute the natural degenerate‑perturbation‑theory basis: each $|\psi_k^{\text{QB}}\rangle$  is an eigenstate of the $\Zthree$ symmetry operator ${\cal P}_{\text{R}1}$ with different eigenvalues. Since [$V_{\text{R1}}, {\cal P}_{\text{R}1}] = 0$, the perturbation cannot mix these states, making them the correct basis choice for evaluating the first‑order perturbative corrections. We take the limit $B \ll \Omega$ to make the perturbative correction to $|\psi_k^{\text{QB}}\rangle$ negligible. The first-order correction yields the following approximate eigenenergies for the cat states: 
\begin{align}
    \epsilon_k &= \bra{\psi_k^{\text{QB}}} V_{\text{R1}}(k) \ket{\psi_k^{\text{QB}}}\label{1-mode-spectrum}\\
    &=2 B e^{-\frac{3}{2}(\lambda/\Omega)^2}\cos\left[ \frac{2 \pi k}{3} + \phi-\frac{\sqrt{3}}{2}\left(\frac{\lambda}{\Omega}\right)^2\right]. \notag
\end{align}
The matrix elements between $\ket*{\psi_k^{\text{QB}}}$ and $\ket*{\psi_l^{\text{QB}}}$ with $k \neq l$ are zero due to the $\Zthree$ symmetry. The comparison of the perturbative result~(\ref{1-mode-spectrum}) with the numerical results obtained by the exact diagonalization of the Hamiltonian~(\ref{simplified-z3-Rabi-Hamiltonian}) is shown in Fig.~\ref{fig:1-mode-rabi}, which demonstrates an excellent agreement between the two. In contrast, the variational method used in Ref.~\cite{sedov_chiral_2020} yields the correct eigenvalues only for $\lambda >  0.61$.

\section{Two-mode Rabi model}
\label{sec:2-mode-rabi}

In this section, we explore several variations of the two-mode $\Zthree$ Rabi model and summarize their key properties. Then, we restrict our attention to the one that is most relevant experimentally. We also show that the latter model exhibits $\Zthree$ cat states in the limit of deep-strong coupling $\lambda \gtrsim \Omega$ and a weak magnetic field $\Omega \gg B$.

\subsection{Hamiltonian of the model}

There are several possible variants of the two-mode $\Zthree$ Rabi model, each featuring a slightly different qutrit-boson interaction. Here, we consider a particular model, whose possible experimental realization has been recently proposed in an optomechanical system~\cite{sedov_chiral_2020}. Furthermore, in our companion paper~\cite{lotkov_superconducting_}, we demonstrate that this model can be realized in a superconducting-circuit platform. The Hamiltonian for this two-mode $\Zthree$ Rabi model is:
\begin{equation}
\label{2-mode-z3-Rabi}
\begin{aligned}
     H_{\text{R}2} = &\Omega ( a_1^{\dagger}  a_1 +  a_2^{\dagger}  a_2) + B (e^{i\phi} Z + e^{-i\phi} Z^{\dagger}) \\
    &- \lambda ( a_1 +  a_2^{\dagger}) X - \lambda ( a_1^{\dagger} +  a_2) X^{\dagger},
\end{aligned}
\end{equation}
where $a_1$ and $a_2$ are canonical boson annihilation operators, $X$ is the $\Zthree$ shift operator, and the parameters $\Omega$, $B$, and $\lambda$ are defined in Eqs.~(\ref{Zn-Rabi-Hamiltonian}) and~(\ref{z3-Rabi-Hamiltonian}).
The Hamiltonian subscript ``R2'' distinguishes the two-mode Rabi model Hamiltonian from the one-mode Rabi model Hamiltonian $H_{\text{R}1}$ in Eq.~(\ref{z3-Rabi-Hamiltonian}).
In Appendix~\ref{app:equivalent_two_mode_Rabi_model}, we show that the model of Ref.~\cite{sedov_chiral_2020} is equivalent to that in Eq.~(\ref{2-mode-z3-Rabi}). 

The model~(\ref{2-mode-z3-Rabi}) is a two-mode extension of the one-mode $\Zthree$ Rabi model~(\ref{z3-Rabi-Hamiltonian}) with the qutrit boson interaction term being different for the two boson modes. Specifically, the qutrit-boson interaction $( a_1 +  a_2^{\dagger}) X + \mathrm{H.c.}$ couples the first mode via its annihilation operator and the second mode via its creation operator. A related two‑mode variant of the one‑mode $\mathbb{Z}_3$ Rabi model $ H_{\text{R}1}$ in Eq.~(\ref{z3-Rabi-Hamiltonian}) with the interaction term $(a_1^{\dagger} + a_2^{\dagger}) X + \mathrm{H.c.}$ is discussed in Appendix~\ref{another-2-mode-z3-Rabi}. However, this two-mode variant turns out to be too similar to the one-mode Rabi model~\eqref{z3-Rabi-Hamiltonian} because they have the same eigenvalues, albeit with different degeneracies.

The $\Zthree$ symmetry generator for the Hamiltonian~\eqref{2-mode-z3-Rabi} has the following form:
\begin{equation}
\label{2-mode-r}
     {\cal P}_{\text{R}2} = \exp\left[\frac{2\pi i}{3} L_3\right] \, Z,\, 
\end{equation}
where $L_3 =  a_1^{\dagger}  a_1 -  a_2^{\dagger}  a_2$ is a generator of the $\mathrm{SU(2)}$ symmetry (see Appendix \ref{su2-symmetry}). Hence, the boson part of the $\Zthree$ symmetry in the two-mode Rabi model~(\ref{2-mode-z3-Rabi}) is a residual part of the $\mathrm{SU(2)}$ symmetry present in the two-dimensional quantum harmonic oscillator.

\subsection{Cat states in the \texorpdfstring{$\Zthree$}{Z\_3} two-mode Rabi model}
\label{subsec:2-mode-spectrum}

To find the spectrum of the two-mode $\Zthree$ Rabi model, we apply the same method we used for the one-mode case. To avoid repetition, we present the derivation steps only briefly. We again introduce new boson operators $ b_1 =  a_1 X$ and $b_2 =  a_2 X^{\dagger}$. After substituting them into the Hamiltonian~\eqref{2-mode-z3-Rabi}, the boson term and the interaction term for each boson mode are transformed into a shifted harmonic oscillator, similarly to Eq.~\eqref{shifted-harmonic-oscillator},
\begin{equation}
\label{2-mode-shifted-harmonic-oscillator}
    \begin{aligned}
\sum_{i=1}^2 & \Omega  a^{\dagger}_i  a_i - \lambda( a^{\dagger}_1 X^{\dagger} +  a_1 X) -\lambda( a^{\dagger}_2 X +  a_2 X^{\dagger}) \\
= \sum_{i=1}^2 &\Omega ( b^{\dagger}_i - \lambda/\Omega)( b_i - \lambda/\Omega) - 2\lambda^2/\Omega.
\end{aligned}
\end{equation}

Then, to completely eliminate the qutrit operator in the Hamiltonian~\eqref{2-mode-z3-Rabi}, we treat the different $\Zthree$ sectors separately. This allows us to express the qutrit clock operator~$Z$ through the new boson operators $b_i$ and $b_i^{\dagger}$ using the definition of the $\Zthree$ symmetry operator~\eqref{2-mode-r}. Consequently, the Hamiltonian~(\ref{2-mode-z3-Rabi}), rewritten in terms of the new boson operators, is naturally divided into an unperturbed part and a perturbation:
\begin{align}
 &H_{\text{R}2}\big|_{\mathcal{H}_k} = H_{\text{R}2}^{(0)}\big|_{\mathcal{H}_k} + V_{\text{R}2}(k)\big|_{\mathcal{H}_k},\, \text{where} \\
 &H_{\text{R}2}^{(0)} = \Omega \sum\limits_{i=1}^2( b_i^{\dagger} - \lambda/\Omega)( b_i - \lambda/\Omega) - 2\lambda^2/\Omega, \\
&V_{\text{R}2}(k)=  2 B \cos\left[\frac{2\pi}{3}( L_3 - k) - \phi\right], \label{2-mode-Rabi-perturbation}
\end{align}
with $\Zthree$ parity taking values $k = 0, \, 1, \, 2$. 

As a result, we have two shifted QHOs with a perturbation depending on the $\Zthree$ parity. The ground states of the unperturbed Hamiltonian in each $\Zthree$ sector are similarly easy to find. They are coherent states of the new boson operator, $b_i\ket*{\psi_k^{\text{Q2B}}} = (\lambda/\Omega)\ket*{\psi_k^{\text{Q2B}}}$, with a fixed $\Zthree$ parity $ {\cal P}_{\text{R}2} \ket*{\psi_k^{\text{Q2B}}} = \omega^k\ket*{\psi_k^{\text{Q2B}}}$. Or, in terms of the original boson and qutrit degrees of freedom the ground states are
\begin{equation}
\label{three-cat-states}
    |\psi_k^{\text{Q2B}}\rangle = \frac{1}{\sqrt{3}}\sum\limits_{l=0}^2 \omega^{lk}|\omega^{-l} \lambda/\Omega\rangle_{\hat a_1}|\omega^l \lambda/\Omega\rangle_{\hat a_2} |\omega^{-l} \rangle,
\end{equation}
where the subscript ``Q2B'' indicates that this is a joint state of a qutrit and two boson modes. Although the states $|\psi_k^{\text{Q2B}}\rangle$ have cat-state structure for any coupling strength, it becomes physically meaningful only once the interaction reaches the deep-strong coupling~\cite{kozin_cavityenhanced_2025,kozin_schottky_2025,vlasiuk_cavityinduced_2023} regime $\lambda \gtrsim \Omega$, where the individual coherent-state components in Eq.~(\ref{three-cat-states}) become well-separated in phase space.

\begin{figure}[t]
\includegraphics[width=\linewidth]{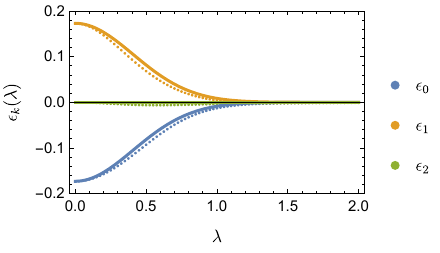}
\caption{Eigenvalues $\epsilon_k$ of the two-mode $\Zthree$ Rabi model $H_{\text{R}2}$ in Eq.~(\ref{2-mode-z3-Rabi}) as functions of coupling $\lambda$ (both are in units of $\Omega$). The dotted line represents the exact numerical diagonalization, while the solid line corresponds to the first-order perturbation theory~(\ref{2-mode-spectrum}). The figure shows that the numerical and the perturbation theory results are very close. Parameters used: $\Omega = 1$, $B = 0.1$, $\phi = 7\pi/6$.}
\label{fig:2-mode-rabi}
\end{figure}

Similarly to the one-mode $\Zthree$ Rabi model, the cat states $|\psi_k^{\text{Q2B}}\rangle$ serve as a unique basis in the ground state of the shifted quantum harmonic oscillator with a specific $\Zthree$ parity: $ {\cal P}_{\text{R}2}|\psi_k^{\text{Q2B}}\rangle = \omega^k |\psi_k^{\text{Q2B}}\rangle$. For later convenience, we denote the subspace spanned by $|\psi_k^{\text{Q2B}}\rangle$ by $\mathcal{R} = \operatorname{Span}[\{|\psi_k^{\text{Q2B}}\rangle\}]$. Because the perturbation~(\ref{2-mode-Rabi-perturbation}) commutes with $ {\cal P}_{\text{R}2}$~\eqref{2-mode-r}, the symmetry prevents mixing between the sectors with different $k$, making $|\psi_k^{\text{Q2B}}\rangle$ the natural basis for the degenerate perturbation theory. 
Just like in Section~\ref{1-mode-perturbation-theory}, we consider the limit $B \ll \Omega$ to make the eigenstate corrections negligible. Using the first-order perturbation theory we find the energy corrections: 
\begin{equation}
\begin{aligned}
\label{2-mode-spectrum}
    \epsilon_k &= \bra{\psi_k^{\text{Q2B}}}V_{\text{R2}}(k) \ket{\psi_k^{\text{Q2B}}} \\
    &= 2 B e^{-3(\lambda/\Omega)^2}\cos\left[\frac{2\pi k}{3} + \phi\right].
\end{aligned}
\end{equation}
The matrix elements between $\ket*{\psi_k^{\text{Q2B}}}$ and $\ket*{\psi_l^{\text{Q2B}}}$ with $k \neq l$ are zero due to the $\Zthree$ symmetry. Note that the cosine phase does not depend on the coherent-state parameter~$\lambda/\Omega$ here. 
The comparison of the perturbative result~(\ref{1-mode-spectrum}) with the numerical results obtained by the exact diagonalization of the Hamiltonian~(\ref{simplified-z3-Rabi-Hamiltonian}) is shown in Fig.~\ref{fig:2-mode-rabi}, which demonstrates an excellent agreement between the two.

One of the reasons why cat states attract significant attention is that they are resilient to phase noise~\cite{mirrahimi_dynamically_2014}. In qubit systems this protection can be extended to bit‑flip errors by switching from the usual two‑component cats to four‑component ones. The same idea carries over to the $\Zthree$ setting: the three‑component cat  states already suppress phase noise, and embedding the logical qutrit in a six‑component cat (as produced, for example, by a $\mathbb{Z}_6$ Rabi model) should also guard against flip errors. 

In this section we showed that the two-mode $\Zthree$ model also has $\Zthree$ cat states as its lowest eigenstates in the deep-strong-coupling regime. 
We now proceed to investigate the properties of these cat states. 

\section{Wigner function} \label{sec:wigner-function}

The Wigner function \cite{wigner_quantum_1932,weyl_quantenmechanik_1927, groenewold_principles_1946} has proven valuable for analyzing cat states in boson systems. At the same time, there is an extensive literature on qudit Wigner functions. The qudit Wigner function based on the finite Heisenberg group is distinct from the $\mathrm{SU(2)}$-based coherent-state construction. The finite Heisenberg group is particularly natural in our context because it already contains the $\mathbb{Z}_n$ symmetry featured throughout this paper. Consequently, to make the structure of the $\mathbb{Z}_3$-symmetric cat states clearer we introduce a Wigner function for a system of a qutrit and two boson modes (Q2B). 

\subsection{Boson Wigner function}

As a reminder, let us briefly review the Wigner function for a single boson mode. In the literature on Wigner functions this system is usually referred to as a continuous-variable (CV) system, and one usually speaks of the CV Wigner function. Here, we use the term boson Wigner function to stay consistent with the rest of the paper.

For a boson mode, we introduce  exponentials of the coordinate and momentum operators, $e^{i p \hat q}$ and $e^{i q \hat p}$ (we use hats to distinguish between parameters $q, p$ and operators $\hat q,\, \hat p$ if it is not clear from the context). From the canonical commutation relation it follows that
\begin{equation}
  \label{boson-commutation-relation}
  e^{i p \hat x} e^{i q \hat p} = e^{iqp} e^{i q \hat p} e^{i p \hat x}.
\end{equation}
These two operators can be interpreted as the coordinate and momentum translation operators respectively.  Combining them yields the displacement operator $ D(q, p)=e^{-iqp/2}\,e^{i p \hat x}e^{-i q \hat p}=e^{z  a^{\dagger}-z^{*} a}$, where we introduced a complex coordinate $z = (q + ip)/\sqrt{2}$. In what follows, we predominantly use complex coordinate $z$ to make the notation more compact. Using the complex coordinate $z$ also simplifies the following statement: the displacement operator $ D(z) $ upon acting on the vacuum creates a coherent state with a coherent-state parameter equal to $z$, i.e., $ D(z)\ket{0}=\ket{z}$. 

For a boson system with a density matrix $\rho$, the Wigner function is defined via the displacement operator as
\begin{equation}
  W_{\text{B}}(\rho;\;z)=\frac{1}{\pi}\Tr\bigl[\rho\, D(2z)\, \Pi\bigr],
\end{equation}
where $\Pi=(-1)^{ a^{\dagger} a}$ is the parity (inversion) operator. In the coordinate basis, it acts as $\Pi\ket{x}=\ket{-x}$.

Since $W_{\text{B}}(\rho; z)$ can take negative values, it is called a \emph{quasi}-probability distribution. While being distinct from the ordinary probability distribution, it retains some of its key properties, i.e., the quasi-probablity distribution is normalized, and its marginal distributions reproduce the correct quantum probabilities,
\begin{gather}
\int \!\mathrm d^2 z\; W_{\text{B}}(\rho;\; z)=1, \\
  \int \!\mathrm d q\; W_{\text{B}}\left(\rho;\; (q + i p)/\sqrt{2} \right)=\bra{p}\rho\ket{p},\\
  \int \!\mathrm d p\; W_{\text{B}}\left(\rho;\; (q +  i p)/\sqrt{2}\right)=\bra{q}\rho\ket{q},
\end{gather}
where we reverted back to $(q,p)$ coordinates for convenience.

\begin{figure}[t]
  \centering
  \subfloat{
    \centering
    \begin{overpic}[width=0.45\linewidth]{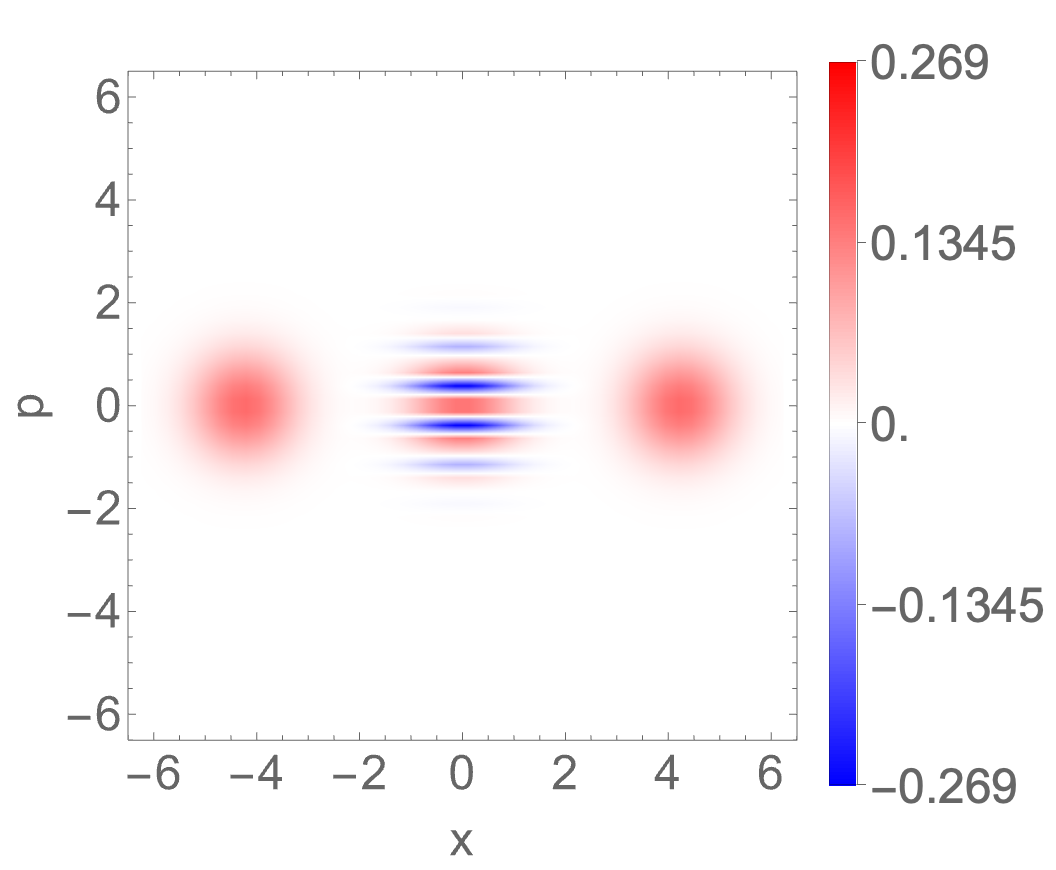}
      \put(1,82){\small (a)} 
    \end{overpic}
    \label{fig:z2-cat-state}
  }
  \subfloat{
    \centering
    \begin{overpic}[width=0.46\linewidth]{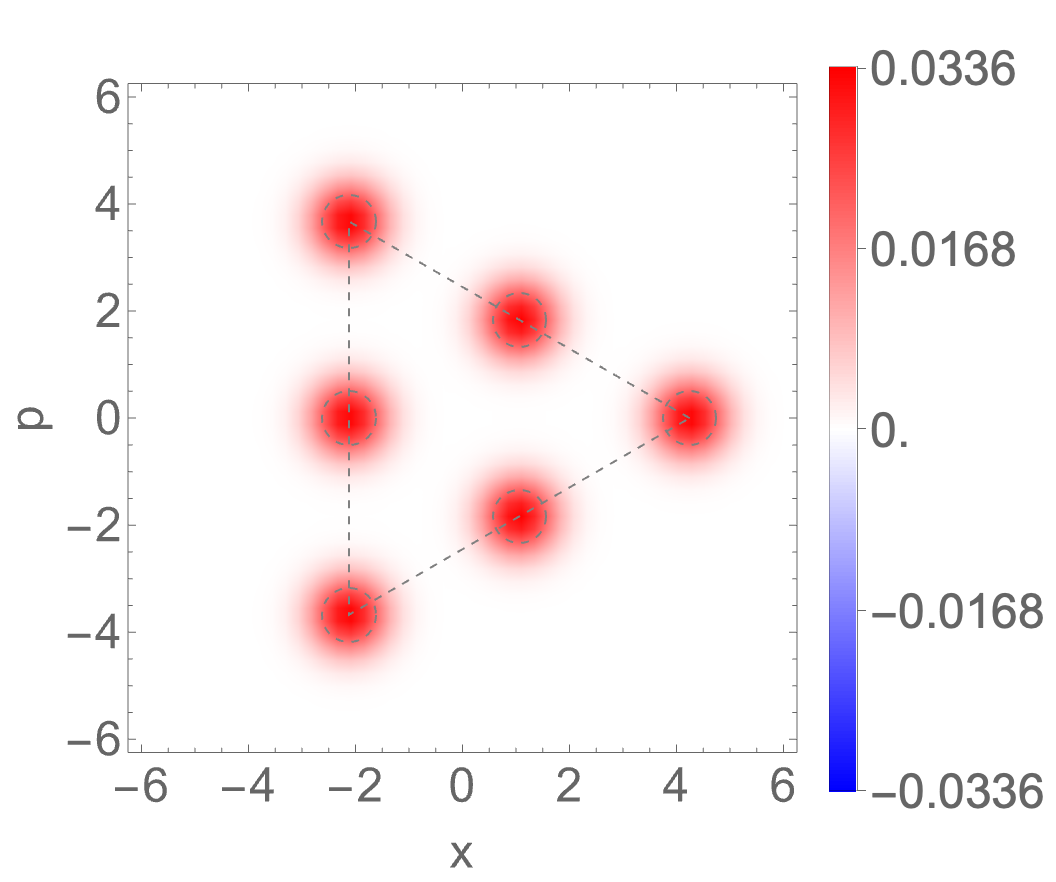}
      \put(1,82){\small (b)}
    \end{overpic}%
    \label{fig:z3-cat-state}
  }
  \caption{Wigner function of the one-boson $\mathbb{Z}_2$ cat state $W_{\text{B}}(\ket{\varphi_{\pm}}\bra{\varphi_{\pm}}; z)$ [given in Eq.~\eqref{z2-cat}] in panel (a); Wigner function of the two-boson $\mathbb{Z}_3$ cat states $W_{\text{2B}}\left(\ket{\psi^{\text{2B}}_k}\bra{\psi^{\text{2B}}_k}; z_1, \, z_2\right)$ [given in Eq.~\eqref{boson-z3-cat-wigner}] in panel~(b). The coherent parameter in both cases is the same, $\alpha = 3$, and we choose to plot the cat state with $k=0$. For the two-boson $\Zthree$ cat state, the Wigner function is four dimensional, and we plot its two-dimensional section along the plane $(z_1 = w/\sqrt{2}, z_2 = w^*/\sqrt{2})$, with $w\in \mathbb{C}$. The dashed triangle is drawn to highlight that all blobs lie on the equilateral triangle. Note that the blobs in the corners of the triangle correspond to the Wigner functions of the three coherent states that comprise the cat state. Meanwhile, the blobs in the middle of the triangle's sides correspond to the interference terms between different coherent states. The fringe pattern, however, is not visible because the oscillations coming from different boson modes compensate each other in the plane $(z_1 = w/\sqrt{2}, \, z_2 = w^*/\sqrt{2})$ with $w \in \mathbb{C}$. The oscillations are still present in the transverse directions (not shown on the plot).}
\end{figure}

Because cat states are central to this work, we recall that the Wigner functions of a purely boson cat states $\ket{\varphi_{\pm}} = (\ket{+\alpha} \pm \ket{-\alpha})/\mathcal{N}_2$ are 
\begin{equation}
  \label{z2-cat}
  \begin{aligned}
  W_{\text{B}}(\ket{\varphi_{\pm}}\bra{\varphi_{\pm}};\; z)=\frac{1}{\pi\mathcal{N}_2}&\left[e^{-2|\alpha - z |^2}+e^{-2|\alpha + z|^2} \right. \\
   \pm 2 &\left.e^{-2|z|^2} \cos\left(4 \Im\{\alpha z^*\} \right)\right],
  \end{aligned}
\end{equation}
where $\mathcal{N}_2$ is a normalization factor and $\alpha$ is a coherent-state parameter of the cat states $\ket{\varphi_{\pm}}$. We do not provide an exact expression for the normalization factor $\mathcal{N}_2$, since for our purposes it is unimportant. The resulting Wigner function is depicted in Fig.~\ref{fig:z2-cat-state}. It consists of two purely exponential terms, those are the Wigner functions of the coherent states $|\pm \alpha\rangle$, and a term with a cosine, which describes the interference between $|+\alpha\rangle $ and $|-\alpha\rangle$.

Taking into account that our final goal is to calculate the Q2B Wigner function for the eigenstates of the two-mode $\mathbb{Z}_3$ Rabi model, which contains one qutrit and two boson degrees of freedom, let us first look at the Wigner function for a two-boson (2B) $\mathbb{Z}_3$ cat state:
\begin{equation}
  \label{2-boson-cat}
    \ket{\psi_k^{\text{2B}}} = \frac{1}{\sqrt{\mathcal{N}_3}}\sum_{j=0}^2 \omega^{jk} |\omega^{-j} \alpha\rangle|\omega^{j}\alpha\rangle .
\end{equation}
The superscript ``2B'' emphasizes that this is a cat state of two bosons. We once again omit an explicit expression for the normalization factor $\mathcal{N}_3$, because it is unimportant. The 2B Wigner function is a simple generalization of the one-boson Wigner function: $W(\rho;\; z_1,z_2) = \frac{1}{\pi^2}\Tr[\rho  D_1(z_1) D(z_2)  \Pi]$, where $D_i(z_i)$ acts only on the $i$th boson, while $\Pi$ acts as the parity operator on both bosons. Applying this definition to $\ket{\psi_k^{\text{2B}}}$, we obtain 
\begin{align}
    &W_{\text{2B}}\left(\ket{\psi^{\text{2B}}_k}\bra{\psi^{\text{2B}}_k};\; z_1, \, z_2\right) \label{boson-z3-cat-wigner}\\
    &= \frac{1}{\mathcal{N}_3} \sum_{i,j=0}^2 \omega^{k(j-i)} W_{\text{B}}\left(|\omega^{-i}\alpha\rangle |\omega^{i}\alpha\rangle \langle \omega^{-j}\alpha|\langle \omega^{j} \alpha|;\; z_1,z_2\right) \notag\\
    &= \frac{1}{\pi^2\mathcal{N}_3} \sum_{b=0}^2 e^{-2 |\alpha - \omega^{b}z_1|^2 - 2|\alpha - \omega^{-b}z_2|^2} \notag\\
    &+  \frac{1}{\pi^2\mathcal{N}_3} \sum_{b=0}^2 e^{- |\alpha + 2\omega^{b}z_1|^2/2 - |\alpha + 2\omega^{-b}z_2|^2/2} \notag\\
    &\times \cos\left[2\sqrt{3}\Re(\alpha[\omega^{b} z_1^* - \omega^{-b}z_2^*]) + 2\pi k/3\right], \notag
\end{align}
where we introduced a new summation index $b = -i - j \mod 3$. In the last expression, the first sum corresponds to the case $i = j$. These terms describe the Wigner functions of the three coherent states $|\omega^{-i} \alpha\rangle |\omega^{i} \alpha \rangle$, while the second sum contains interference terms between the different coherent states $|\omega^{-i} \alpha\rangle |\omega^{i} \alpha \rangle$ and $|\omega^{-j} \alpha\rangle |\omega^{j} \alpha \rangle$, with $i \neq j$. The latter are a counterpart of the fringe-pattern term for the ordinary cat state~(\ref{z2-cat}). 
All six terms in Eq.~(\ref{boson-z3-cat-wigner}) are four-dimensional Gaussians, whose centers all lie on the plane $(z_1 = w\alpha/(\sqrt{2}|\alpha|), \, z_2 = w^*\alpha/(\sqrt{2}|\alpha|))$, with $w \in \mathbb{C}$ and form a triangle [see Fig.~\ref{fig:z3-cat-state}].

The same recipe is sometimes applied to hybrid qudit-boson (or qubit-boson) cat states. One first takes the partial trace over the finite-dimensional subsystem and then plots the boson Wigner function of the reduced state. This procedure, however, discards all information about the qudit-boson entanglement. We therefore advocate using the full qudit-boson Wigner function instead.

\subsection{Qutrit Wigner function}

We now explain how the Wigner functions can be applied to qutrit systems. The qutrit shift ($X$) and clock ($Z$) operators are analogous to the boson translation operators $e^{ip\hat q}$ and $e^{-iq\hat p}$, respectively.
Indeed, the boson operators satisfy the commutation relation~\eqref{boson-commutation-relation}, and the qutrit operators fulfill the same algebra~$ZX=\omega XZ$.
Consequently, the qutrit displacement operators can be introduced in a manner similar to the boson Wigner function: $D_{\text{Q}}(a,b)= \omega^{-ab/2} Z^b X^a$. Therefore, the qutrit Wigner functions can be defined as
\begin{equation}
  W_{\text{Q}}(\rho;\; q,p) = \frac{1}{3} \Tr[\rho D_{\text{Q}}(2q,2p) \Pi_{\text{Q}}],
\end{equation}
where $\rho$ is a qutrit density matrix and $\Pi_{\text{Q}}$ is the qutrit parity operator, i.e., $\Pi_{\text{Q}} |i\rangle = |-i\rangle$, with $i = 0,1,2$.
Here $q$ and $p$ are discrete phase-space coordinates taking values $0,1,2$, which correspond to the eigenvalues of the $Z$ and $X$ operators, respectively.

The qudit Wigner functions have been extensively studied~\cite{wootters_wignerfunction_1987,luis_discrete_1998,leonhardt_discrete_1996,gibbons_discrete_2004,zak_doubling_2011}. Notably, the Wigner function definition for even- and odd-dimensional qudits differs significantly, although a unifying framework has been recently proposed~\cite{antonopoulos_grand_2025}. For our purposes such generality is unnecessary, and we restrict ourselves only to the qutrit case. However, all our results can be easily generalized to arbitrary odd-dimensional qudits. While a similar construction is also possible for the even-dimensional case, it is considerably more involved.

These Wigner functions share the basic properties of their boson counterparts. They are normalized, and their marginal distributions reproduce the physical probabilities,
\begin{gather}
  \sum_{q,p=0}^2 W_{\text{Q}}(\rho;\; q,p)= 1, \\
  \sum_{q=0}^2 W_{\text{Q}}(\rho;\; q,p)=\bra{\omega^{p}}\rho \ket{\omega^{p}},\\
  \sum_{p=0}^2 W_{\text{Q}}(\rho;\; q,p)=\bra{q}\rho\ket{q}.
\end{gather}

As a simple illustration, the Wigner functions of the computational basis states (i.e., the eigenstates of $Z$) are given by $W_{\text{Q}}\left(\ket{i}\!\bra{i};\; q,p\right) =(1/3)\,\delta_{q,i}$. Similarly, for the eigenstates of $X$ one has $W_{\text{Q}}\left(\ket{\omega^{j}}\!\bra{\omega^{j}};\; q,p\right) = (1/3)\,\delta_{p,j}$.
Thus, we see that the qutrit Wigner function yields the expected quasi-probability distributions.

\subsection{Qutrit-boson Wigner function}

Finally, we have all necessary ingredients to discuss a joint {\it qudit-boson} (QB) Wigner function. While the construction itself is straightforward, one only has to combine the qudit Wigner function and the boson Wigner function, we believe that the QB Wigner function is a powerful tool to study the hybrid systems. In this subsection, we first calculate the QB Wigner function for the Q2B cat state. Then, we present the numerical calculations of the QB Wigner functions for the two-mode $\Zthree$ Rabi model, which clearly demonstrates the formation of Q2B cat states in the deep-strong-coupling regime.

\subsubsection{Analytical derivation in the deep-strong-coupling regime}

\begin{figure*}
  \centering
  \includegraphics[width=\linewidth]{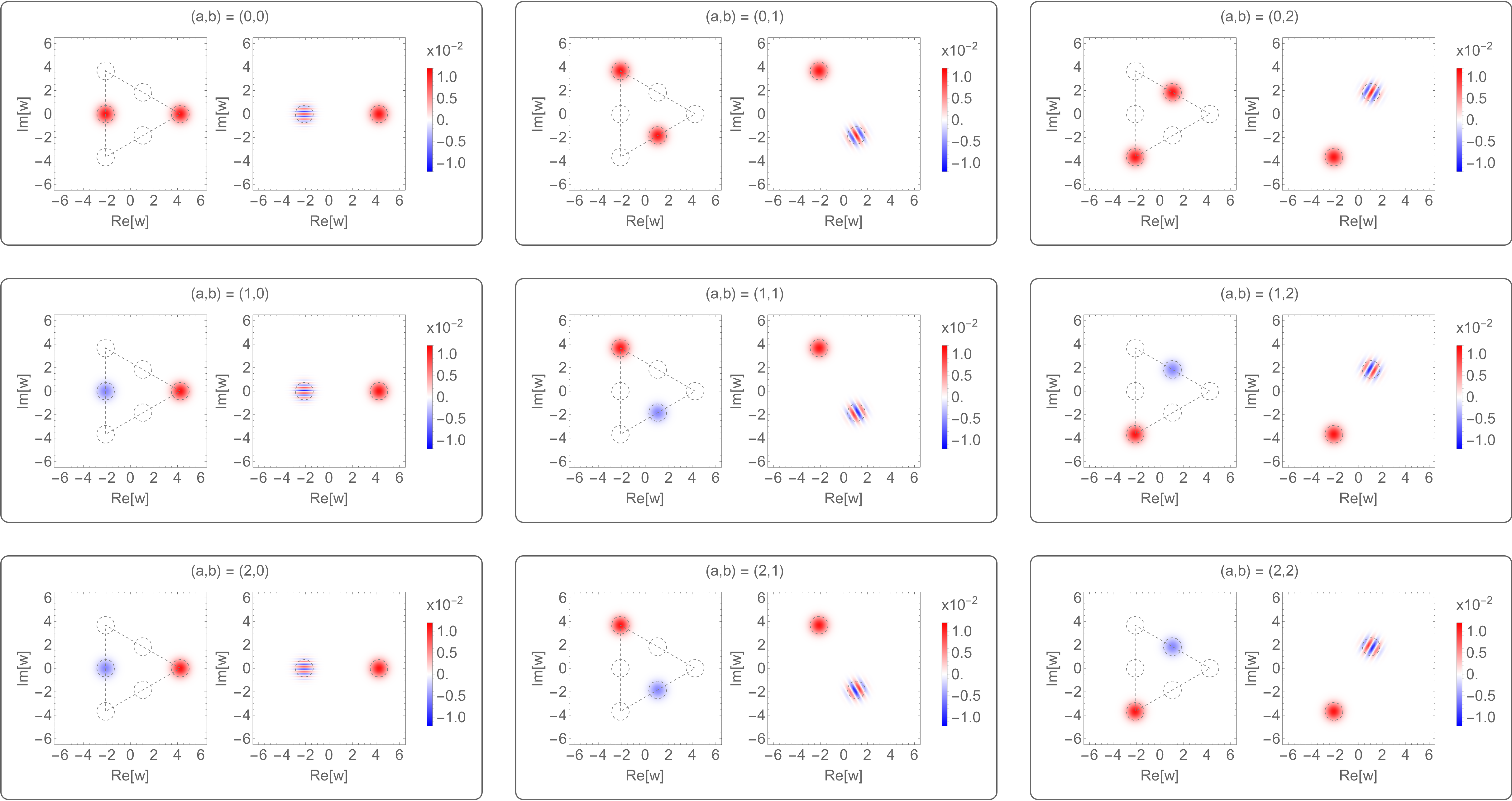}
  \caption{Qutrit-boson Wigner function of the qutrit-boson cat state $W_{\text{Q2B}}\left(\ket*{\psi_k^{\text{Q2B}}}\bra*{\psi_k^{\text{Q2B}}}; z_1,z_2, a,b\right)$ [given by Eq.~\eqref{joint-qb-cat-wigner}] with the coherent-state parameter $\alpha = 3$, the plotted cat state number is $k=0$. The $3\times 3$ grid allows us to plot the two discrete qutrit variables $(a,b)$. For each qurtrit variables values $(a,b)$, we show two plots of the qutrit-boson Wigner function along two different sections of the four-dimensional space $(z_1,z_2),\, z_1,z_2 \in \mathbb{C}$. The left plot in each pair shows the four-dimensional two-boson Wigner functions along the two-dimensional plane $(z_1 = w/\sqrt{2}, \, z_2 = w^*/\sqrt{2}),\, w\in \mathbb{C}$. The section is the same for all $(a,b)$ because all non-zero blobs are located on this plane similarly to the purely boson case. The right plot in each pair shows the four-dimensional two-boson Wigner functions along the two-dimensional plane $\left(z_1 = \omega^b \{\cos(\pi/4 + 2\pi b/3) \Re(w) + \sin(\pi/4 + 2\pi b/3) \Im(w)\}, \, z_2 = \omega^{-b} \{\cos(\pi/4 - 2\pi b/3) \Re(w) - \sin(\pi/4 - 2\pi b/3) \Im(w)\}\right)$, $w\in \mathbb{C}$. This plane was chosen since it simultaneously shows both blobs present in $(a,b)$ part of the qutrit-boson Wigner function and the oscillating fringes of the interference term. The dashed circles schematically shows the Wigner function of the 2B cat state $\ket{\psi_k^{\text{2B}}}$~\eqref{boson-z3-cat-wigner} in the same section for a comparison. The interference terms between different coherent states correspond to the blobs in the middles of the triangle's sides.}
  \label{fig:joint-qb-wigner}
\end{figure*}

Since we are now dealing with both qutrit and boson degrees of freedom, it is convenient to define a joint displacement operator
\begin{equation}
    D_{\text{Q2B}}(z_1, z_2, a, b) = D_{\text{B1}}(z_1) D_{\text{B2}}(z_2) D_{\text{Q}},
\end{equation}
where $D_{\text{B}i}$, with $i =1,2$ is a displacement operator for the $i$th boson mode and $D_{\text{Q}}$ is a displacement operator for the qutrit.
Similarly, one can introduce a joint parity operator
\begin{equation}
    \Pi_{Q2B} = \Pi_{B1} \Pi_{B2} \Pi_Q
\end{equation}
where $\Pi_{\text{B}i}$ with $i=1,2$ is the parity operator for the $i$th boson mode and $\Pi_{\text{Q}}$ is the qutrit parity operator.
Then, the Q2B Wigner function of a qutrit-boson density matrix $\rho_{\text{Q2B}}$ can be defined in the standard way
\begin{equation} \label{q2b-wigner-fun-general}
  \begin{aligned}
      &W_{\text{Q2B}}(\rho_{\text{Q2B}};\; z_1, z_2, a, b) \\
      &= \frac{1}{3\pi}\Tr\left[\rho_{\text{Q2B}} D_{\text{Q2B}}(2z_1,2z_2,2a, 2b)\Pi_{\text{Q2B}}\right].
  \end{aligned}
\end{equation}

In the deep-strong-coupling regime, the eigenstates of the $\Zthree$ Rabi model are the $\Zthree$ cat states given by Eq.~\eqref{three-cat-states}. Substituting them into the Q2B Wigner function~(\ref{q2b-wigner-fun-general}) yields
\begin{align}
    &W_{\text{Q2B}}\left(\ket{\psi_k^{\text{Q2B}}}\bra{\psi_k^{\text{Q2B}}};\; z_1,z_2, a,b\right) \label{joint-qb-cat-wigner}\\
    &= \frac{1}{9\pi^2}\left\{e^{-2|\alpha - \omega^{-b} z_1|^2 - 2|\alpha - \omega^b z_2|^2} + \right. \notag \\
    &+ e^{-|\alpha + 2\omega^{-b} z_1|^2/2 - |\alpha + 2 \omega^{b} z_2|^2/2} \notag\\&\times\left. \cos\left[2\sqrt{3}\Re(\alpha[\omega^{b} z_1^* - \omega^{-b}z_2^*]) + 2\pi(a+k)/3\right] \right\}, \notag
\end{align}
where $\alpha = \lambda/\Omega$ is a coherent state parameter. 
The Q2B Wigner function~(\ref{joint-qb-cat-wigner}) is shown in Fig.~\ref{fig:joint-qb-wigner}.

By comparing Eqs.~(\ref{boson-z3-cat-wigner}) and (\ref{joint-qb-cat-wigner}), we see that the Wigner function of the 2B cat states~(\ref{boson-z3-cat-wigner}) (see Fig.~\ref{fig:z3-cat-state}) has six terms, whereas the Wigner function of the Q2B cat states~(\ref{joint-qb-cat-wigner}) (see Fig.~\ref{fig:joint-qb-wigner}) has only two terms.
It is to be expected because in the Q2B $\Zthree$ cat state the boson modes are entangled with the qutrit. As a result, the six terms of the 2B $\Zthree$ cat state are spread over different values of the qutrit variable~$b$. Namely, the plots corresponding to $(a=0, b=0)$ consist of the coherent state Wigner function $|\alpha\rangle|\alpha\rangle$ and the Wigner function of the interference term between the states $|\omega\alpha\rangle|\omega^{2}\alpha\rangle$ and $|\omega^{2}\alpha\rangle|\omega\alpha\rangle$. In a similar manner for other values of $b$, the Wigner function $W_{\text{Q2B}}(\rho_{\text{Q2B}};\; z_1, z_2, a=0, b)$ consists of a coherent state term $|\omega^{2b}\alpha\rangle$, located at the corners of the triangle in Fig.~\ref{fig:joint-qb-wigner} and an interference term between the remaining two coherent states, located at the midpoints of the triangle's sides. As a result, if we sum the Q2B Wigner function~(\ref{joint-qb-cat-wigner}) over $b$ while keeping $a = 0$, we will obtain the result identical to Eq.~(\ref{boson-z3-cat-wigner}). 

Next, we discuss the Q2B Wigner function plot of the QB cat state~\eqref{joint-qb-cat-wigner} shown in Fig.~\ref{fig:joint-qb-wigner} more thoroughly. First of all, since the Q2B Wigner function depends on the discrete variables $a,b \in \{ 0,\, 1, \, 2\}$, Fig.~\ref{fig:joint-qb-wigner} shows nine pairs of plots labeled by $(a,b)$ instead of a single one. The left-hand plot in a pair corresponding to a specific $(a,b)$ values shows the Q2B Wigner function~\eqref{joint-qb-cat-wigner} along the two-dimensional section defined by $(z_1 = w/\sqrt{2}, \, z_2 = w^*/\sqrt{2}),\, w\in \mathbb{C}$. In other words, it is a plot of $W_{\text{Q2B}}\left(\ket*{\psi_k^{\text{Q2B}}}\bra*{\psi_k^{\text{Q2B}}};\; w/\sqrt{2},w^*/\sqrt{2}, a,b\right)$ as a function of $w$. Because all the exponents in Eq.~\eqref{joint-qb-cat-wigner} are centered on this plane, the left plots have the same section for all $(a,b)$ values. Unfortunately, the fringes of the interference term are not visible along this plane because the two bosons' oscillations perfectly compensate each other. On the other hand, the section plane of the right-hand plot explicitly depends on the value of the qutrit variable $b$. Specifically, the section plane of the $(a,b)$ right-hand plot is 
\begin{equation}
    \begin{aligned}
        &\left(\omega^b \{\cos\!\left(\frac{\pi}{4} + \frac{2\pi b}{3} \right)\Re(w) + \sin\!\left(\frac{\pi}{4} + \frac{2\pi b}{3}\right) \Im(w)\},\right. \\
        &\left.\, \omega^{-b} \{\cos\!\left(\frac{\pi}{4} - \frac{2\pi b}{3} \right)\Re(w) -\! \sin\left(\frac{\pi}{4} - \frac{2\pi b}{3}\right) \Im(w)\}\right),
    \end{aligned}
\end{equation}
with $w\in \mathbb{C}$. This plane is chosen to show the fringes of the interference term since they are not visible on the left plot.

Both the qutrit variable $a$ and the Q2B cat state's number $k$ enter the Wigner function~\eqref{joint-qb-cat-wigner} together through the cosine phase as $2\pi(a+k)/3$. Consequently, the Q2B Wigner function plots for the cat states $\ket*{\psi^{\text{QB}}_k}$ with different $k = 0,1,2$ differ only by a cyclic permutation of rows in Fig.~\ref{fig:joint-qb-wigner}. Moreover, as changing the value of $a$ with fixed $k$  only affects the phase of the fringe oscillation, inspecting the $a=0$ row of the QB Wigner function suffices to verify the Q2B-cat structure. The $a=0$ Q2B Wigner function of the cat state~\eqref{joint-qb-cat-wigner} already shows that the qutrit and two bosons are entangled with each other: the six blobs are evenly distributed among the three plot pairs, and the right-hand plots clearly display the fringe oscillations. The $a\neq 0$ plots show the same picture only with a different phase of the fringe oscillation. Therefore, below we plot only three out of the nine plots, $(a=0, b)$.

\subsubsection{Visualizing entanglement in hybrid qutrit-boson systems}

One of the common applications of the ordinary Wigner functions is to distinguish between cat states and maximally mixed states. The Q2B Wigner function generalizes this to the hybrid Q2B system. First, consider the ordinary boson system: the cat state~$\rho_{\text{cat}}$ and the maximally-mixed state~$\rho_{\text{mix}}$ are given by
\begin{align}
        \rho_{\text{cat}} &= \frac{1}{\mathcal{N}_2}\biggl(|+\alpha\rangle \pm |-\alpha\rangle\biggr)\biggl(\langle +\alpha | \pm \langle -\alpha|\biggr), \\ 
        \rho_{\text{mix}} &= \frac{1}{2}\biggl(|+\alpha\rangle\langle +\alpha| + |-\alpha\rangle\langle -\alpha|\biggr).
\end{align}
The Wigner function can easily discern between the two. If the fringe pattern is present, then it is a cat state. Otherwise, it is a maximally-mixed one. 

\begin{figure}[t]
  \centering
  \subfloat{
    \centering
    \begin{overpic}[width=0.9\linewidth]{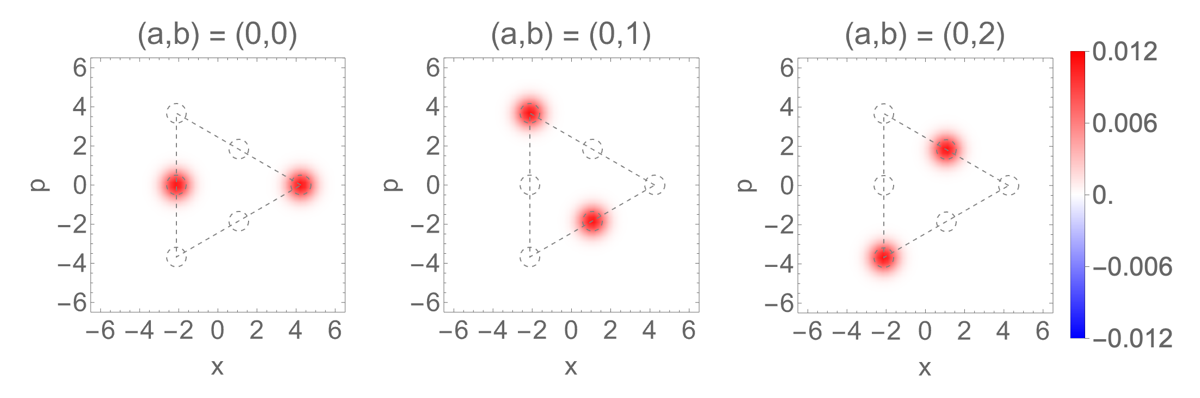}
      \put(1,30){\small (a)} 
    \end{overpic}
    \label{fig:q2b-cat}
  }\\
  \subfloat{
    \centering
    \begin{overpic}[width=0.9\linewidth]{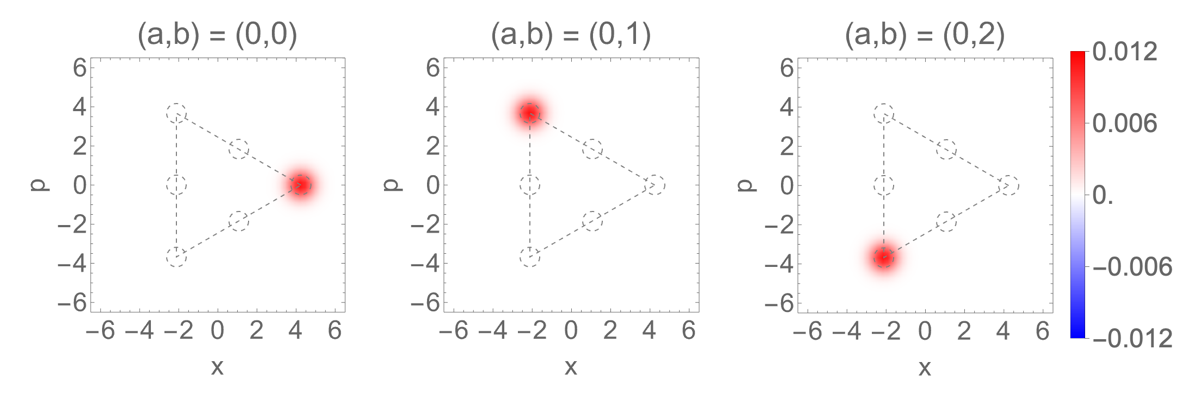}
      \put(1,30){\small (b)} 
    \end{overpic}
    \label{fig:mixed-cat}
  }\\
  \subfloat{
    \centering
    \begin{overpic}[width=0.9\linewidth]{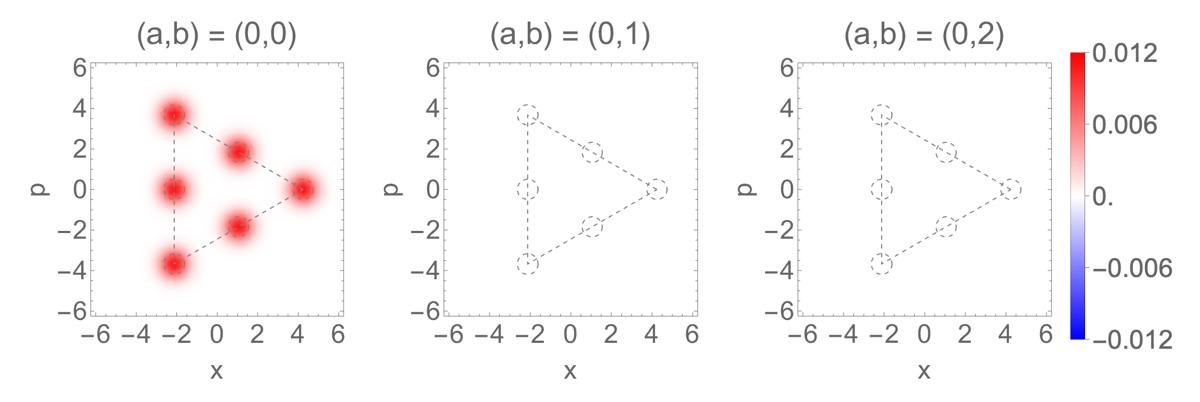}
      \put(1,30){\small (c)} 
    \end{overpic}
    \label{fig:2b-cat}
  }\\
  \caption{Comparison of the Q2B Wigner functions $W_{\text{Q2B}}$ [given by Eq.~\eqref{q2b-wigner-fun-general}] for three states: (a) Q2B cat state $\rho_{\text{Q2B-cat}}$ with $k = 0$  defined in Eq.~\eqref{q2b-density-matrix}, (b) maximally mixed state $\rho_{\text{mix}}$ in Eq.~\eqref{mixed-cat-density-matrix}, and (c) 2B cat state unentangled with the qutrit $\rho_{\text{2B-cat}}$ in Eq.~\eqref{2b-cat-density-matrix}. The coherent-state parameter is chosen to be the same for all three states, $\alpha = 3$. For each state the three plots $(a=0, b = 0,1,2)$ are shown. With the two-boson-mode Wigner function being four dimensional, we show its section along the two-dimensional plane $(z_1 = w/\sqrt{2}, \, z_2 = w^*/\sqrt{2})$ with $ w\in \mathbb{C}$ since all non-zero regions are centered on this plane. The dashed triangle schematically shows the Wigner function of the 2B cat state~\eqref{boson-z3-cat-wigner} to enhance the plots readability.}
  \label{fig:cats-comparison}
\end{figure}

Conceptually, for the Q2B systems the situation is very similar. However, the number of different options is larger.
The Q2B Wigner function~(\ref{joint-qb-cat-wigner}) allows us to distinguish between all possible cases. Specifically, the three plots with $(a=0, b=0,1,2)$ look drastically different for a Q2B cat state, a maximally-mixed state, and a 2B cat state unentangled with the qutrit:
\begin{itemize}
    \item[({\it i}\,)] The Q2B Wigner function plots of the {\it Q2B cat state},
    \begin{equation}
    \label{q2b-density-matrix}
        \rho_{\text{Q2B-cat}} = \ket{\psi_0^{\text{Q2B}}}\bra{\psi_0^{\text{Q2B}}},
    \end{equation}
    are shown in Fig.~\ref{fig:q2b-cat}. Every plot has two blobs: one in the triangle's corner, i.e., the coherent state Wigner function, and one in the midpoint of the opposite side, i.e., the Wigner function of the interference term.
    \item[({\it ii}\,)] The Q2B Wigner function plots of the {\it maximally-mixed state},
    \begin{equation}
    \label{mixed-cat-density-matrix}
        \rho_{\text{mix}} = \frac{1}{3}\sum_{i=0}^2\ket{\omega^{-i}\alpha}\ket{\omega^{i}\alpha}\ket{\omega^{i}}\bra{\omega^{-i}\alpha}\bra{\omega^{i}\alpha}{\bra{\omega^{i}}},
    \end{equation}
    are shown in Fig.~\ref{fig:mixed-cat}. Every plot has a single blob in the triangle's corner, i.e., the Wigner function for the coherent state. The blobs corresponding to the interference terms are absent since $\rho_{\text{mix}}$ is a classical mixture of the three coherent states. 
    \item[({\it iii}\,)] The Q2B Wigner function plots of the {\it 2B cat state~\eqref{2-boson-cat} unentangled with the qutrit},
    \begin{equation}
    \label{2b-cat-density-matrix}
        \rho_{\text{2B-cat}} = \ket{\psi_k^{\text{2B}}}\ket{\omega^0}\bra{\psi_k^{\text{2B}}}\bra{\omega^0},
    \end{equation}
    are shown in Fig.~\ref{fig:2b-cat}. Here, the $(a=0, b=0)$ plot shows the Wigner function of the 2B cat state (identical to the one in Fig.~\ref{fig:z3-cat-state}), while the rest of the plots are empty. It shows that the boson modes and the qutrit are not entangled. 
\end{itemize}

As one can see, the Q2B Wigner function clearly distinguishes between the three states. Therefore, it proves to be a powerful tool to study the hybrid qudit-boson system akin to the ordinary Wigner function in the purely boson systems.

\subsubsection{Numerical calculation outside of the deep-strong-coupling regime}

\begin{figure*}[t]
  \centering
  \subfloat{
    \centering
    \begin{overpic}[width=0.9\linewidth]{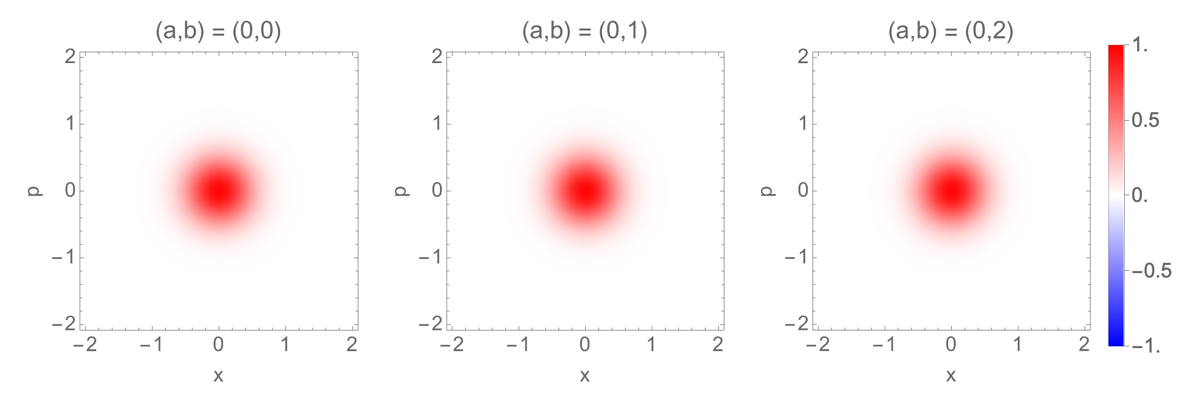}
    \put(1,32){\small (a)}
    \end{overpic}
    \label{fig:numerical-wigner-l01}
  }\\
  \subfloat{
    \centering
    \begin{overpic}[width=0.9\linewidth]{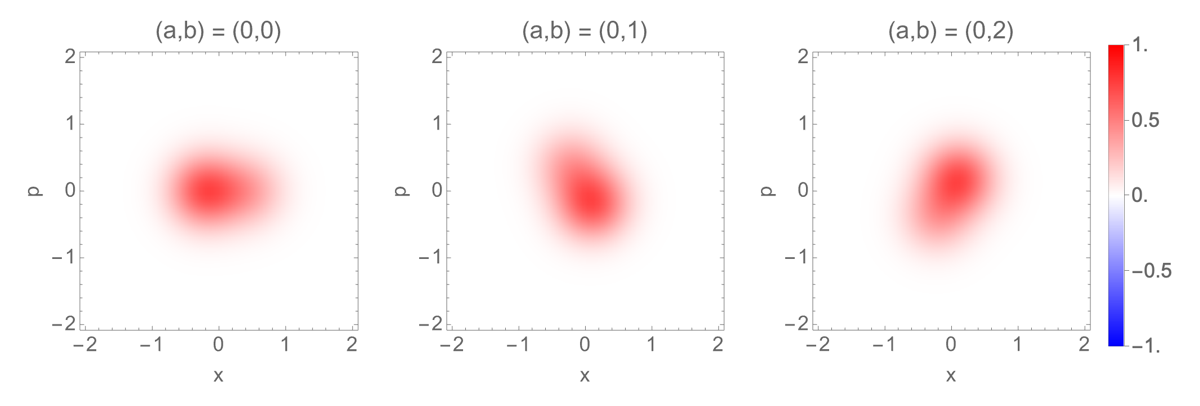}
    \put(1,32){\small (b)}
    \end{overpic}
    \label{fig:numerical-wigner-l05}
  }\\
  \subfloat{
    \centering
    \begin{overpic}[width=0.9\linewidth]{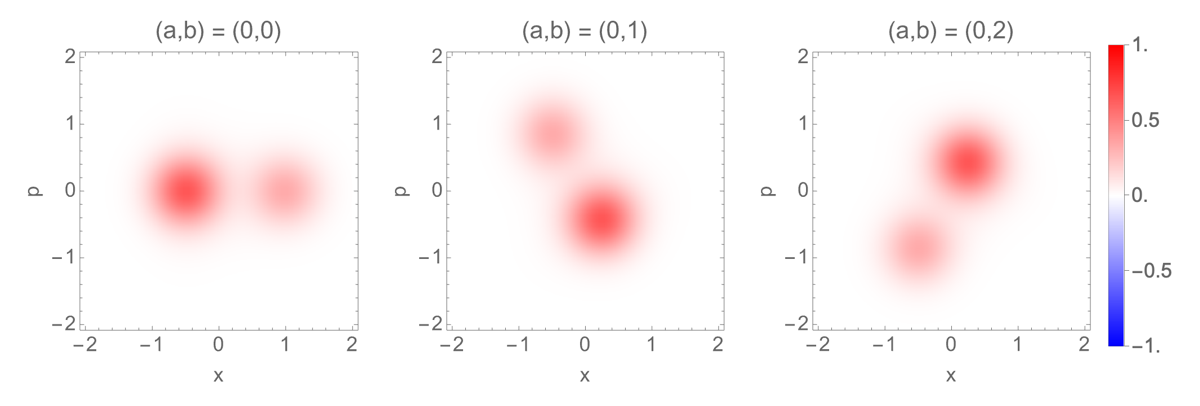}
    \put(1,32){\small (c)}
    \end{overpic}
    \label{fig:numerical-wigner-l1}
  }\\
  \caption{Numerically-calculated qutrit-boson Wigner function for the ground state of the two-mode $\Zthree$ Rabi model $H_{\text{R2}}$ defined in Eq.~\eqref{2-mode-z3-Rabi} with coupling constants: (a) $\lambda/\Omega = 0.1,\, B/\Omega = 0.1$, (b) $\lambda/\Omega = 0.5,\, B/\Omega = 0.1$, and (c) $\lambda/\Omega = 1,\, B/\Omega = 0.1$. To save space, we plot only the first row ($a = 0$) of the $3\times 3$ grid of the qutrit-boson Wigner function for each value of the coupling constant $\lambda/\Omega$. This is sufficient to see the cat-state structure emerging when the coupling strength increases. The four-dimensional two-boson Wigner functions are plotted along the two-dimensional plane $(z_1 = w, \, z_2 = w^*),\, w\in \mathbb{C}$. For the numerical calculations we truncated the Hilbert space of both boson modes at $\ket{n=50}$ in the occupational basis.}
  \label{fig:numerical-wigner}
\end{figure*}

Now that we have introduced and discussed the Wigner functions for the hybrid Q2B system, we calculate them numerically for the eigenstates of the two-mode $\mathbb{Z}_3$ Rabi model outside of the deep-strong-coupling regime, where analytical results are not available. Our results clearly demonstrate how the $\Zthree$ cat states emerge on approach into the deep-strong-coupling regime ($\lambda \gtrsim \Omega$ and $\Omega \gg B$).
In Fig.~\ref{fig:numerical-wigner} we plot the first row ($a = 0$) of the Q2B Wigner function for three different coupling constants $ \lambda/\Omega = 0.1,\, 0.5, \, 1$.  Similarly to the behavior of the ordinary cat states, the blobs move apart when the coherent parameter $\alpha = \lambda/\Omega$ is increased. At $\lambda/\Omega = 1$ they become well-separated. In Fig.~\ref{fig:numerical-wigner}, once again  we demonstrate what we have already proved analytically in Sec.~\ref{sec:2-mode-rabi}: the three lowest eigenstates of the two-mode $\mathbb{Z}_3$ Rabi model become the Q2B $\mathbb{Z}_3$ cat states in the limit $\lambda \gtrsim \Omega$ and $\Omega \gg B$.

\section{Conclusions} \label{sec:conclusions}

We have studied one-mode [given by Eq.~\eqref{z3-Rabi-Hamiltonian}] and two-mode [given by Eq.~\eqref{2-mode-z3-Rabi}] variations of  the $\Zthree$ quantum Rabi model. We proposed a canonical transformation that allowed us to find the $\Zthree$ Rabi model's spectrum in the limit of deep-strong coupling. In this limit, the three lowest eigenstates are $\Zthree$ cat states. Then we studied these states using joint qutrit-boson Wigner functions. In particular, we derived an explicit expression for the Wigner function of the qutrit-boson cat state. Also, we calculated the Wigner function of the $\Zthree$ Rabi model's eigenstates numerically for different coupling strengths.

With the $\Zn$ Rabi model recently attracting significant attention, our results should further the understanding of these models, in particular, their similarities and differences with the conventional $\Ztwo$ Rabi model. Moreover, the proposed approach based on the joint qutrit-boson Wigner function improves on the ordinary Wigner function allowing one to study the entanglement between the qutrit and boson subsystems. The approach readily generalizes to qubit-boson systems and thus promises to be generally useful for studies of the light-matter interaction.

While in this paper we extensively studied the regime of the deep-strong coupling and weak magnetic field, the understanding of the $\Zn$ Rabi model in other coupling regimes remains an open problem. Compared to the ordinary Rabi model, various parameter regimes are still to be studied. This provides a direction for future work.

\section*{Acknowledgments} 

We thank Daria Kalacheva, Henry Legg, Katharina Laubscher, and Ilia Luchnikov for fruitful discussions and useful comments. This work was supported as a part of NCCR SPIN, a National Centre of Competence in Research, funded by the Swiss National Science Foundation (grant number 225153). This work has received funding from the Swiss State Secretariat for Education, Research and Innovation (SERI) under contract number M822.00078.

\appendix

\section{\texorpdfstring{Equivalence between the Hamiltonian~(\ref{2-mode-z3-Rabi}) and the model of Ref.~\cite{sedov_chiral_2020}}
        {Equivalence between the two-mode Z\_3 Rabi Hamiltonian  
         and the Sedov et al. model}}
\label{app:equivalent_two_mode_Rabi_model}

The two-mode $\Zthree$ Rabi model~(\ref{2-mode-z3-Rabi}) is equivalent to the model considered in Ref.~\cite{sedov_chiral_2020}. The latter reads
\begin{equation}
\label{coordinate-2-mode-z3-Rabi}
\begin{aligned}
     H_{\text{R}2'} = &\Omega ( a_1^{\dagger}  a_1 +  a_2^{\dagger}  a_2) + B(e^{i\phi} Z + e^{-i\phi} Z^{\dagger}) \\
    &- \lambda ( x_1 + i  x_2)  X - \lambda ( x_1 - i  x_2) X^{\dagger},
\end{aligned}
\end{equation}
where we defined $x_{1,2} = (a_{1,2} + a_{1,2}^{\dagger})/\sqrt{2}$ and $p_{1,2} = (a_{1,2} - a_{1,2}^{\dagger})/(\sqrt{2}i)$. In this appendix, $x_i$ and $p_i$ with $i=1,2$ denote the canonical coordinate and momentum operators as opposed to the classical coordinates in Sec.~\ref{sec:wigner-function}.

This representation of the two-mode $\Zthree$ Rabi model can be transformed into our Hamiltonian $ H_{\text{R}2}$ in Eq.~(\ref{2-mode-z3-Rabi}) by a canonical transformation that mixes the two boson modes.
The $\Zthree$ symmetry generator in this case takes the form: 
\begin{equation} \label{z3-symmetry-generator-equivalent}
  {\cal P}_{\text{R}2'} = \exp[\frac{2\pi i}{3} (x_1  p_2 -  x_2  p_1)] Z,
\end{equation}
 Note that here the $\Zthree$ symmetry is realized in a peculiar way. The boson part of the $\Zthree$ symmetry generator~(\ref{z3-symmetry-generator-equivalent}) is just a discrete rotation in the plane of the two-dimensional quantum harmonic oscillator.

\section{Another version of the two-mode \texorpdfstring{$\Zthree$}{Z3} Rabi model}
\label{another-2-mode-z3-Rabi}

In this appendix, we examine another two-mode extension of the $\Zthree$ Rabi model~(\ref{z3-Rabi-Hamiltonian}), which is more straightforward compared to the one used in the main text [see Eq.~(\ref{2-mode-z3-Rabi})]. Here we simply double the number of boson modes without altering the Hamiltonian~(\ref{z3-Rabi-Hamiltonian}) in any other way:
\begin{equation}
\label{another-2-mode-z3-Rabi-hamiltonian}
\begin{aligned}
     H = &\Omega ( a_1^{\dagger}  a_1 +  a_2^{\dagger}  a_2) + B (e^{i\phi} Z + e^{-i\phi} Z^{\dagger}) \\
    &- \lambda \left[( a_1^{\dagger} +  a_2^{\dagger}) X + ( a_1 +  a_2) X^{\dagger}\right].
\end{aligned}
\end{equation}

Correspondingly, the $\Zthree$ symmetry generator is
\begin{equation}
     {\cal P} = \exp[\frac{2\pi i}{3} N] \, Z,\, \text{with}\  N =  a_1^{\dagger}  a_1 +  a_2^{\dagger}  a_2.
\end{equation}

Interestingly, unlike the two-mode $\Zthree$ Rabi model~(\ref{2-mode-z3-Rabi}) discussed in the main text, the Hamiltonian~(\ref{another-2-mode-z3-Rabi-hamiltonian}) possesses the $\mathrm{SU(2)}$ symmetry of the 2D QHO. As a result, in this system we have the $ \Zthree $ and $\mathrm{SU(2)}$ symmetries combined into $ G = \Zthree \times \mathrm{SU(2)}$. This leads to a degeneracy of the spectrum similar to the one we saw in Appendix~\ref{su2-symmetry}. In this sense, the addition of the second mode did not change anything but the multiplicities of the eigenenergies. As a consequence, we find the two-mode $\Zthree$ Rabi model discussed in Sec. \ref{sec:2-mode-rabi} to be more interesting compared to the one in Eq.~(\ref{another-2-mode-z3-Rabi-hamiltonian}).

\section{\texorpdfstring{$\mathrm{SU(2)}$}{SU(2)} symmetry in two-dimensional quantum harmonic oscillator}
\label{su2-symmetry}

\begin{figure}[t]
    \centering
    \includegraphics[width=\linewidth]{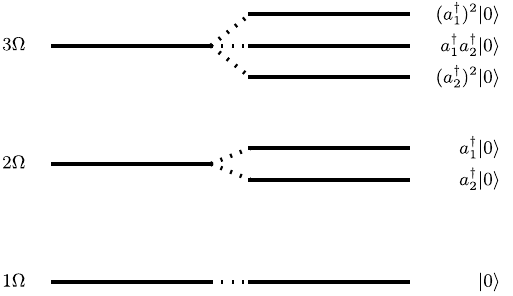}
    \caption{Eigenvalue degeneracy of two-dimensional quantum harmonic oscillator with the Hamiltonian~\eqref{2d-qho}.}
    \label{fig:2d-qho-diagram}
\end{figure}

While the two-dimensional quantum harmonic oscillator (2D QHO) is discussed in any introductory quantum mechanics course, for the sake of completeness we would like to briefly review a certain fact relevant to our purposes. The Hamiltonian of the 2D QHO is
\begin{equation}
\label{2d-qho}
  H = \Omega ( a_1^{\dagger}  a_1 +  a_2^{\dagger}  a_2 + 1).
\end{equation}

Keeping in mind that the one-dimensional QHO eigenenergies are simply $\Omega(n+1/2)$, $n = 0,1,2,3,\dots$, we can easily deduce the spectrum of the 2D QHO. Here, we have two modes, in which we can put excitations. As a result, the $n$th eigenenergy, $n \geq 1$, becomes degenerate. Using the basic combinatorical fact that there are $(n+1)$ ways to distribute $n$ objects onto two groups, it follows that the $n$th energy level has $(n+1)$-fold degeneracy. This looks quite similar to the $\mathrm{SU(2)}$ group representations, i.e., there is exactly one $\mathrm{SU(2)}$ representation on a $n$-dimensional vector space. This is not a coincidence: the 2D QHO has a $\mathrm{SU(2)}$ symmetry which ``rotates'' the excitations between two modes.  

The explicit expressions for the $\mathrm{SU(2)}$ symmetry generators are given by:
\begin{equation}
\begin{aligned}
 L_1 &=  a_1^{\dagger}  a_2+  a_2^{\dagger}  a_1, \\
 L_2 &= i  a_1^{\dagger}  a_2-i  a_2^{\dagger}  a_1, \\
 L_3 &=  a_1^{\dagger}  a_1 -  a_2^{\dagger}  a_2.
\end{aligned}
\end{equation}

Despite this being a simple and well-known fact, the presence of the $\mathrm{SU(2)}$ symmetry is relevant for our discussion of the two-mode $\Zthree$ Rabi model, which is a 2D QHO coupled with a three-level system (qutrit). The interaction between the 2D QHO and the $3$-level system can either break the $\mathrm{SU(2)}$ symmetry or preserve it.

\bibliography{references}

\end{document}